\documentclass[twocolumn,showpacs,amsmath,aps]{revtex4}
\usepackage{graphicx,color}
\usepackage{CJK}
\usepackage{bm}
\usepackage[hypertex]{hyperref}

\newcommand{\be}{\begin{equation}}
\newcommand{\ee}{\end{equation}}
\newcommand{\bea}{\begin{eqnarray}}
\newcommand{\eea}{\end{eqnarray}}
\newcommand{\bsube}{\begin{subequations}}
\newcommand{\esube}{\end{subequations}}

\newcommand{\Eq}[1]{Eq.\,(\ref{#1})}
\newcommand{\Eqs}[1]{Eqs.\,(\ref{#1})}

\newcommand{\dg}{\dagger}
\newcommand{\la}{\langle}
\newcommand{\ra}{\rangle}

\newcommand{\nl}{\nonumber \\}

\newcommand{\beq}{\begin{equation}}
\newcommand{\eeq}{\end{equation}}
\newcommand{\beqn}{\begin{eqnarray}}
\newcommand{\eeqn}{\end{eqnarray}}
\newcommand{\bsub}{\begin{subequations}}
\newcommand{\esub}{\end{subequations}}

\begin{document}

\title{ Quantum trajectories under frequent measurements
        in non-Markovian environment }

\author{Luting Xu}
\email{xuluting@bnu.edu.cn}
\affiliation{Center for Advanced Quantum Studies and
Department of Physics, Beijing Normal University,
Beijing 100875, China}
\author{Xin-Qi Li}
\email{lixinqi@bnu.edu.cn}
\affiliation{Center for Advanced Quantum Studies and
Department of Physics, Beijing Normal University,
Beijing 100875, China}

\date{\today}

\begin{abstract}
In this work we generalize the quantum trajectory (QT)
theory from Markovian to non-Markovian environments.
We model the non-Markovian environment
by using a Lorentzian spectral density function
with bandwidth ($\Lambda$),
and find perfect ``scaling" property
with the measurement frequency ($\tau^{-1}$)
in terms of the scaling variable $x=\Lambda\tau$.
Our result bridges the gap
between the existing QT theory and the Zeno effect,
by rendering them as two extremes corresponding to
$x\to\infty$ and $x\to 0$, respectively.
This $x$-dependent criterion improves the idea
of using $\tau$ alone, and quantitatively identifies
the validity condition of the conventional QT theory.
\end{abstract}

\pacs{03.65.Ta,03.65.Xp,73.63.-b,73.40.Gk}
\maketitle

{\flushleft The quantum trajectory (QT)}
given by stochastic Schr\"odinger equation (SSE)
for an open system associated with {\it Markovian dynamics}
can be interpreted as quantum state conditioned on
continuous observation (monitoring) on the environment
\cite{Dali92,WM93}.
The QT theory of this type has been well demonstrated
and broadly applied \cite{WM09,Jac14}, including the
recent experiments in superconducting solid-state circuits
\cite{Pala10,DiCa13,Hof11,Mar11,Dev13,Sid13,Sid12,DiCa12,Dev13a}.
On the other hand,
associated with the {\it non-Markovian dynamics} of open quantum systems,
similar non-Markovian stochastic Schr\"odinger equation (nMSSE)
has been constructed \cite{Str98,Str99}.
However, the nMSSE is largely
a working tool of unraveling the non-Markovian dynamics,
which cannot be interpreted as measurement-conditioned
{\it physical} quantum trajectory \cite{Wis08,Dio08,Wis03}.
After careful analysis by Wiseman {\it et al},
the nMSSE might be at most interpreted as certain
``hidden variable" theory, i.e., taking the complex Wiener variable
$z_t$ involved in the nMSSE as an ``objective property"
which inherently exists in the environment,
rather than a consequence of continuous measurements \cite{Wis03}.

In this work we consider the interesting problem
how to construct the {\it physical} QT associated with
frequent monitoring on non-Markovian environment.
To be specific, we model the non-Markovian environment
by using a Lorenztian spectral density function (SDF)
with finite bandwidth.
We show that the result is quite different
from the nMSSE mentioned above.
Elegantly, via slight modification by involving a ``scaling" variable,
the resultant QT formally resembles, but essentially generalizes,
the conventional QT.
Our result bridges the gap between the existing QT
\cite{Dali92,WM93,WM09,Jac14}
and the quantum Zeno effect \cite{Zeno-1},
by rendering them as two extremes
which have quite different predictions \cite{SG13,SG14}.

Let us consider a two-level atom (qubit) prepared
in a quantum superposition of the ground state ($|g\ra$)
and exited state ($|e\ra$),
$|\Psi(0)\ra =\alpha_0|e\ra + \beta_0|g\ra$.
Now consider its evolution under
{\it continuous} (very frequent) measurements in the
surrounding environment for the spontaneous emission of photon.
From the celebrated QT theory \cite{Dali92,WM93,WM09,Jac14},
 conditioned on the {\it continuous} null-result
(no-register of spontaneous emission) detection,
the state would change, following the simple formula
\bea\label{NR-1}
|\Psi(t)\ra = \left( \alpha_0 e^{-\Gamma t/2} |e\ra
+ \beta_0|g\ra \right) /\, {\cal N} \,,
\eea
where $\Gamma$ is the spontaneous emission rate
and ${\cal N}$ denotes the normalization factor.
To interpret this result, reasoning based on
{\it informational} evolution is sometimes put forward.
That is, {\it no result} is a sort of {\it information},
so the state can change according to Bayesian inferring,
similar as in classical probability theory.

On the other hand, the above continuous null-result
quantum motion is prohibited by the quantum Zeno effect \cite{Zeno-1}.
We may briefly summarize the treatment and result as follows.
Starting with $|\Psi(0)\ra$, let us expand the evolution operator
up to the second order in $\tau$, $U(\tau)\simeq 1-iH\tau-H^2\tau^2/2$,
where $\tau$ is the time interval between the successive
null-result measurements. Each null-result measurement
would project the wave function on the atomic subspace.
Consider $n$ subsequent null-result measurements during time $t$
(with $n=t/\tau$). In the limit $\tau\to 0$ and $t$=const,
one obtains (see Appendix A for more details)
\begin{align}\label{ZE-1}
|\Psi_n\rangle \to \alpha_0|e\rangle
+\beta_0|g\rangle \equiv |\Psi (0)\rangle   \, .
\end{align}
So we find that the frequent null-result monitoring of the environment
will prevent the change of the state,
resulting thus in the quantum Zeno effect.

Actually, the QT theory leading to \Eq{NR-1}
is from unraveling the Markovian Lindblad master equation.
In Markovian approximation, one requires a wide bandwidth
environment (i.e., the bandwidth $\Lambda\to\infty$).
Therefore, any $\tau$ is long compared to the environment's
memory time $\Lambda^{-1}$, leading thus to the exponential
decay of population which destroys the possibility of Zeno effect.
In the case of $\Lambda\to\infty$, the above expansion on
$U(\tau)$ is invalid.
In order to generate Zeno effect, the physical condition
is $\tau << \Lambda^{-1}$.
In the remainder of this work, we will develop a treatment
to smoothly bridge these two extremes,
and construct the associated QT theory
by introducing external drive to the atom.

\vspace{0.2cm}

{\flushleft\it Spontaneous decay}. ---
The two-level atom coupled to the
electromagnetic vacuum (environment) is described by the Hamiltonian
\begin{align}
H= \frac{\Delta_{eg}}{2}\sigma_z
+\sum_r\left(b^{\dg}_{r}b_r+\frac{1}{2}\right)\omega_r
+\sum_{r} \left[V_{r} b^{\dg}_{r}\sigma^{-}
     + {\rm H.c.} \right] \, .
\end{align}
Throughout this work we set $\hbar=1$.
Here we introduce: the two-level energy difference
$\Delta_{eg}=E_e-E_g$,
the atomic operators
$\sigma_z=|e\ra \la e|-|g\ra \la g|$, $\sigma^-=|g\ra \la e|$,
and $\sigma^+=|e\ra \la g|$. $V_{r}$ is the
coupling amplitude of the atom with the environment.
Then, consider the evolution of the entire system,
starting with an initial state
$|\Psi(0)\rangle=(\alpha_0|e\rangle+\beta_0|g\rangle)
\otimes |{\rm vac}\ra$,
where $|{\rm vac}\ra$ stands for the environmental vacuum
with no photon. Under the influence of the coupling,
the entire state at time $t$ can be written as
\bea\label{WF-1}
|\Psi(t)\rangle &=& \alpha(t)|e\rangle\otimes |{\rm vac}\ra
+ \sum_r c_r(t)|g\rangle\otimes |1_r;0;{\cdots}\ra  \nl
&& + \, \beta_0|g\rangle\otimes |{\rm vac}\ra  \,,
\eea
where $|1_r;0;{\cdots}\ra$ describes the environment
with a photon excitation in the state ``$r$"
and no excitations of other states.
The coefficients have initial conditions of
$\alpha(0)=\alpha_0$ and $c_r(0)=0$.

Substituting Eq.~(\ref{WF-1}) into the Schr\"odinger equation
and performing the Laplace transform, one can obtain
the solution of $\alpha(t)$ in frequency domain
(see Appendix B for more details).
That is, replace $\sum_r |V_r|^2 [\cdots]
\to\int d\omega_r D(\omega_r)[\cdots]$,
where $D(\omega_r)=\sum_{r'}|V_{r'}|^2
\delta(\omega_r-\omega_{r'})
\to D_0\Lambda^2/[ (\omega_r-\omega_0)^2+\Lambda^2] $
is the spectral density function (SDF),
approximated here by a finite-band Lorentzian spectrum with $\omega_0$
the spectral center and $\Lambda$ the width \cite{Kof05}.
We obtain then the time-dependent amplitude
$\alpha(t)\equiv a(t)\alpha_0$
via the inverse Laplace transform as \cite{SG14}
\begin{align}
a(t)={1\over A_+^{}-A_-^{}}(A_+^{}e^{-A_-^{}t}-A_-^{}e^{-A_+^{}t}) \,,
\label{proj0}
\end{align}
with $A_{\pm}^{}=[\Lambda -iE\pm
\sqrt{(\Lambda -iE)^2-2\Gamma\Lambda}]/2$.
Here we introduced the energy offset
$E=(E_e-E_g)-\omega_0$
and the usual decay rate in wide-band limit, $\Gamma=2\pi D_0$.

\vspace{0.2cm}
{\flushleft\it Frequent null-result measurements}.---
The null-result measurement in the environment,
quantum mechanically,
collapses the entire wave function onto the atomic subspace.
After $n$ such null-result measurements
with subsequent time interval $\tau=t/n$,
the final state of the atom is
\begin{align}
|\widetilde{\Psi}(t)\ra =
\left[\bar{a}(t)\alpha_0|e\ra+\beta_0|g\ra \right] /\sqrt{{\cal N}_n(t)} \,,
\end{align}
where $\bar{a}(t)=a^n(\tau)$ and
${\cal N}_n(t)=|\bar{a}(t)\alpha_0|^2+|\beta_0|^2$.
Note that, unlike the case of the wide-band-limit Markovian environment,
$|\widetilde{\Psi}(t)\ra$
differs from the single-null-measurement-collapsed
state at the final moment from $|\Psi(t)\ra$.
It can be proved that the normalization factor ${\cal N}_n$
equals also the {\it joint} probability of getting
{\it null} results in all the intermediate measurements,
i.e., $(1-\sum_r|c_r(\tau)|^2)^n$.
Let us denote ${\cal N}_n(t)\equiv p_{0}^{(n)}(t)$.
Accordingly, during time $(0,t)$,
the probability of detecting a spontaneous photon
is $p_{1}^{(n)}(t)=1-p_{0}^{(n)}(t)$.

\begin{figure}
\includegraphics[scale=0.55]{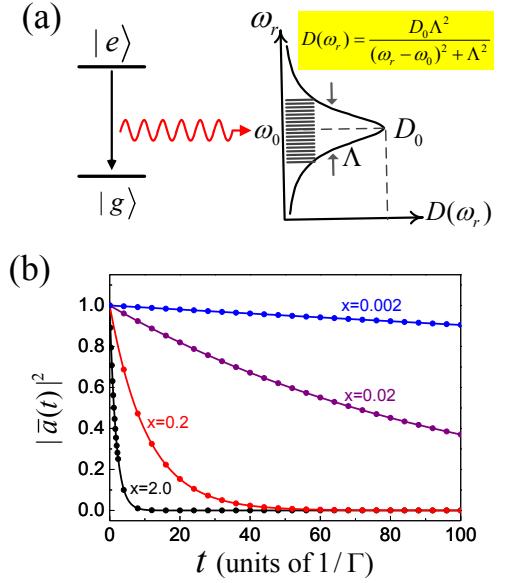}
\caption{ (Color online)
(a)
Spontaneous emission of a two-level atom coupled to
non-Markovian environment with finite-bandwidth Lorentzian spectrum.
(b)
Effective decay factor of the excited state
started with a quantum superposition
$\alpha_0|e\ra+\beta_0|g\ra$,
under frequent null-result measurements in the environment.
Scaling behavior is demonstrated by the remarkable agreement
between \Eq{scal} (continuous lines)
and $a^n(\tau)$ (symbols) calculated using \Eq{proj0}
with $\Lambda=10\Gamma$ (as an example) and $E=\omega-\omega_0=0$.
Remind that $t=n\tau$ and $x=\Lambda\tau$.   }
\end{figure}

Now let us consider the limit of ``continuous" measurements,
$n\to\infty$ by taking the measurement time interval
$\tau\to 0$ and keeping $t=n\tau$ fixed.
Supposing to increase the bandwidth $\Lambda$
so that the variable $x=\Lambda\tau$ remains constant,
we can prove a ``scaling" property
that the final state becomes a function of $x$ only.
To reveal the full scaling behavior in general case,
we also assume the energy offset $E=c\Lambda$
(in usual treatment $c=0$).
One finds from Eq.~(\ref{proj0}) that
$A_+^{}= \kappa\Lambda -\Gamma/(2\kappa)$
and $A_-^{}=\Gamma/(2\kappa)$
(up to the order of $(\Gamma/\Lambda )^2$),
where $\kappa=1-ic$. Using
$\big(1-{z\over n}\big)^n=e^{-z\big(1+{z\over 2n}+\cdots\big)}$
and neglecting small terms $\sim \Gamma/\Lambda$ in exponent,
we arrive to \cite{SG14}
\begin{align}
\bar a(t)=a^n_{}(\tau)=\exp\left\{
-\left[\frac{1}{\kappa}-( 1-e^{-\kappa x})\frac{1}{\kappa^2x}
\right]\frac{\Gamma t}{2}  \right\}   \,.
\label{scal}
\end{align}
Elegantly, this result reveals an explicit scaling property
in the $x=\Lambda \tau$-variable.
In Fig.\ 1(b), by relaxing the conditions
($n\to\infty$ and $\tau\to 0$) for obtaining this analytic formula,
we illustrate the scaling behavior in broad parameter conditions.

\vspace{0.2cm}
{\flushleft\it Some remarks about \Eq{scal}}.---
{\it (i)}
The numerical results in Fig.\ 1(b) for finite $\Lambda$ and $\tau$
(e.g., the ``$x=2$" curve for and $\tau^{-1}=0.5\Lambda$)
show excellent agreement with \Eq{scal},
indicating that we can expect the ``scaling" behavior
beyond the limits $n\to\infty$ and $\tau\to 0$.
This limiting procedure is only a mathematical technique
leading us to obtain the analytic result \Eq{scal}.

{\it (ii)}
The scaling behavior can be understood
via the time-energy uncertainty relation.
Actually, the successive measurements with time interval $\tau$
in the reservoir will cause fluctuations
of the atom's level ($E_e$) by amount $\sim \tau^{-1}$,
since the result
whether or not a spontaneous emission detected in the reservoir
allows knowing whether or not the atom is in the excited state.
Then, if we (conceptually) expand the width of the reservoir's SDF
by this same amount (i.e., by $\sim \tau^{-1}$),
we can expect the same (identical) decay dynamics.
This is the physical reason of the scaling behavior
shown analytically by Eq.\ (7) and numerically in Fig.\ 1(b).

{\it (iii)}
Note that the $x$-dependence of the decay dynamics
is the same as the $\tau$-dependence
for a given bandwidth $\Lambda$
(usually it is difficult to change $\Lambda$ in real set-ups).
And, this $\tau$-dependence is the essential feature
associated with measurements in non-Markovian reservoir,
which is in sharp contrast with the conventional
{\it $\tau$-independent} Markovian case.

{\it (iv)}
From \Eq{scal}, in the wide-band limit, $x\to\infty$ and $\kappa\to 1$,
one recovers the result $\bar a(t)\to e^{-\Gamma t/2}$
predicted by the standard QT theory.
On the other hand, in the limit of $x\to 0$,
one finds from \Eq{scal} that $\bar a(t)=1$,
so that the atom is frozen in its initial state,
showing the Zeno effect.

{\it (v)}
In the Zeno regime $\tau^{-1}>>\Lambda$, one may encounter
a ``negative frequency" problem if the central frequency
$\omega_0$ is not much larger than $\Lambda$.
In this case (and for the transition energy $\Delta_{eg}>\omega_0$)
the level $E_e$ may fluctuate into
the domain of ``negative frequency" of the SDF, thus violating
the condition of the symmetric Lorenztian SDF model
and needing certain modification to \Eq{scal}.
In this work, we assume a {\it symmetric} Lorentzian SDF model
under the conditions $\Delta_{eg}>\omega_0>>\Lambda$,
for the sake of showing
a full transition from the Markovian behavior to Zeno effect
governed by the unified \Eq{scal}.
In this case, there is no ``negative frequency" difficulty
to affect the validity of \Eq{scal}.

{\it (vi)}
From \Eq{scal}, one can define an {\it effective} decay rate
\bea\label{rate-1}
\gamma_{\rm eff}={\rm Re}\left\{\left[ 1- (\kappa x)^{-1}
\left( 1-e^{-\kappa x} \right) \right]/\kappa \right\}\, \Gamma  \,.
\eea
Note that for the wide-band-limit Markovian environment
the exponential decay process implies {\it no-effect}
of the intermediate null-result interruptions \cite{SG14}.
\Eq{rate-1}, however, shows that
the decay rate is influenced by the frequent null-result measurements.
This $x$- or $\tau$-dependence reflects the non-Markovian effect
rooted in \Eq{proj0}, despite that the frequent measurements
cut off the usual non-Markovian correlation (memory) effect
between different $\tau$-period evolutions.
It is right the accumulation of the ``small" non-Markovian contributions
over $t=n\tau$ that makes \Eqs{scal} and (\ref{rate-1})
and the associated QT (to be constructed)
generalize the usual Markovian results.

\vspace{0.2cm}
{\flushleft\it Quantum trajectories}.---
Corresponding to direct photon detection, let us first
construct the Monte-Carlo wave function (MCWF) approach,
closely along the line proposed in Ref.\ \cite{Dali92}.
Consider the state evolution under frequent null-result measurements
between $t$ and $t+\Delta t$, with thus $\Delta t=n\tau$.
The probability with
photon register in the detector during $\Delta t$,
is $p^{(n)}_{1}(\Delta t)=|\alpha(t)|^2 \gamma_{\rm eff}\Delta t$.
Under the ``scaling" consideration,
the effective decay rate $\gamma_{\rm eff}$
is simply given by \Eq{rate-1}, or, alternatively by
\bea\label{rate-2}
\gamma_{\rm eff}= [1-|\bar{a}(\Delta t)|^2]/\Delta t
~~{\rm or}~~
\gamma_{\rm eff}= -\ln[\,|\bar{a}(\Delta t)|^2]/\Delta t  \,.  \nl
\eea
For small $\Delta t$, which implies $|\bar{a}(\Delta t)|^2\simeq 1$,
both definitions are equivalent and coincide with \Eq{rate-1}.

In practical simulations,
generate a random number $\epsilon$ between 0 and 1.
If $\epsilon < p^{(n)}_{1}(\Delta t)$,
which corresponds to the probability of having
a photon register in the detector ($\Delta N_c=1$),
we update the state by a ``jump" action
\bea
|\widetilde{\Psi}(t+\Delta t)\ra= \sigma^- |\widetilde{\Psi}(t)\ra
/\parallel\bullet \parallel  \,,
\eea
where $\parallel\bullet \parallel$ denotes the normalization factor.
On the other hand, if $\epsilon > p^{(n)}_{1}(\Delta t)$,
which corresponds to the NRM with $\Delta N_c=0$,
we update the state via the effective {\it smooth} evolution
\bea
|\widetilde{\Psi}(t+\Delta t)\ra
= {\cal U}(\Delta t) |\widetilde{\Psi}(t)\ra
/\parallel\bullet \parallel   \,.
\eea
In terms of a matrix form defined by
$\{\alpha(t+\Delta t),\beta(t+\Delta t)\}^T
= {\cal U}(\Delta t)\{\alpha(t),\beta(t)\}^T$,
the effective {\it non-unitary} evolution operator reads
\begin{align}
{\cal U}(\Delta t)=\left(
\begin{array}{cc}
\bar{a}(\Delta t) & 0 \\[5pt]
0 & 1
\end{array}    \right) \,.
\end{align}
Noting that $\Delta t=n\tau$, as above,
here we mention again that
$\bar{a}(\Delta t)=[a(\tau)]^n$ which can be \Eq{scal}
in the limit $\tau\to 0$ and $n\to \infty$,
or be more generally determined using \Eq{proj0} for $a(\tau)$.


Based on the MCWF approach proposed above,
one can simulate the (stochastic) quantum trajectories
under frequent photon detections in the environment.
Ensemble average over these trajectories of quantum (pure) state
corresponds to the result given by the following master equation
\cite{Dali92,WM93,WM09,Jac14}:
\bea\label{ME-1}
\dot{\rho}=-i[H_S,\rho]
+ \gamma_{\rm eff} {\cal D}[\sigma^-]\rho \,,
\eea
where ${\cal D}[\bullet]\rho\equiv
\bullet\rho \bullet^{\dg}
-\frac{1}{2}\{\bullet^{\dg}\bullet,\rho \}$.
Formally, this is an $x$- or $\tau$-dependent
Lindblad-type master equation.
However, unlike its Markovian counterpart,
a significant difference lies in the fact that
this equation does {\it not} describe
the {\it reduced} state $\varrho(t)$ of the (open) quantum system.
It is well known that $\varrho(t)$ is defined by
tracing the environment degrees of freedom
from the entire (system-plus-environment)
unitary wavefunction at time $t$.
Here, ``tracing" simply means performing projective measurements
and making average {\it only} at the last moment $t$,
on the entire unitary wavefunction evolved
from the same {\it initial} state.
In contrast to $\varrho(t)$,
the state $\rho(t)$ given by \Eq{ME-1}
is the ensemble-averaged state of
the system under successive measurement interruptions.
Remarkably, the successive measurements would destroy
the correlation effect
between {\it different} $\tau$-period evolutions,
resulting thus in the Markovian-Lindblad-type \Eq{ME-1}
with, however, an effective $\gamma_{\rm eff}$
rather than certain ``natural" decay rate.


Following Refs.\ \cite{Dali92,WM93,WM09,Jac14},
we now include external driving into \Eq{ME-1},
via $H_S=\frac{\Delta_{eg}}{2}\sigma_z+\Omega\sigma_x$.
Note that the validity of this procedure is rooted
in the {\it additivity of the state changes}
over the very small time interval ($\tau$).
As a result,
there are two contributions to the state change:
one is {\it informational} owing to the continuous
measurements, and the other is {\it physical}
which is caused by the external driving.
Note also that in general the dissipative two-level atom
under driving is not exactly solvable.
The underlying complexity can be imagined as follows:
there are more and more photons emitted into the reservoir;
and the emitted photon can re-excite the atom.
However, {\it in the presence of frequent measurements},
the emitted photon will be destroyed by detectors.
During each successive measurement interval ($\tau$),
it is reasonable to assume that
there is {\it at most} one photon in the reservoir.
Therefore, even in the presence of external driving,
\Eq{ME-1} is valid under the above considerations.

Instead of the {\it direct} detection of the spontaneous emission
considered above, one can also adopt the so-called homodyne
detection scheme by mixing the emitting photons
with a classical field with modulating phase $\varphi$ \cite{WM93,WM09}.
The measurement result (optical current) of this type
can be expressed as \cite{WM93,WM09},
$ I_{\varphi}(t)=\sqrt{\gamma_{\rm eff}}
\la \sigma^- e^{-i\varphi} + \sigma^+ e^{i\varphi}\ra /2 + \xi(t) $,
where $\la\cdots \ra={\rm Tr}[(\cdots)\rho(t)]$
and $\xi(t)$ is the Gaussian white noise associated with quantum jumps.
In this measurement scheme, the detection result
is a sum of the classical reference field
and the photon emitted by the atom.
The ``jump" (knowledge change of atom state)
associated with photon register
in the detector is relatively weak,
developing thus a ``diffusive" regime
because of the mix of the reference field.
Through a careful analysis \cite{WM93,WM09},
the difference of the detected result (in single realization)
during $(t,t+dt)$ from the expected one using earlier $\rho(t)$,
is characterized by $\xi(t)dt$ in the expression of $I_{\varphi}(t)$.
Conditioned on $I_{\varphi}(t)$, the state evolution is
given by the {\it diffusive} QT equation \cite{WM93,WM09}:
\bea\label{QTE}
\dot{\rho}=-i[H_S,\rho]
+ \gamma_{\rm eff} {\cal D}[\sigma^-]\rho
+ \sqrt{\gamma_{\rm eff}} {\cal H}[e^{-i\varphi}\sigma^-]\rho \xi(t),
\eea
where ${\cal H}[\bullet]\rho\equiv \bullet\rho+\rho \bullet^{\dg}
-\la \bullet+\bullet^{\dg}\ra \rho $.
Essentially, \Eq{QTE} generalizes the existing QT equation by
accounting for the measurement frequency ($\nu=1/\tau$)
in the effective ``spontaneous" emission
rate $\gamma_{\rm eff}$.

\begin{figure}
\includegraphics[scale=0.75]{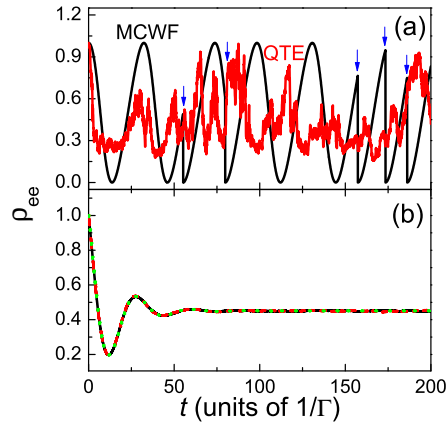}
\caption{ (Color online)
(a) Two quantum trajectories
from the MCWF (black) and QT equation (red) simulations.
The blue arrows indicate quantum ``jumps" owing to
``direct" detection of the spontaneous emission of the atom.
(b)
Ensemble average of 2000 MCWF and QT equation trajectories
and the result (green curve) from the master equation
$\dot{\rho}=-i[H_S,\rho]
+ \gamma_{\rm eff} {\cal D}[\sigma^-]\rho$.
Parameters used in the simulation:
$\Omega=0.1$, $\Gamma=1.0$, $x=0.2$ and $E=0$.   }
\end{figure}

In Fig.\ 2(a) we display two representative quantum trajectories
from the MCWF and the diffusive QT equation (\ref{QTE}).
We see that the former type of quantum trajectory reveals
drastic ``quantum jump" owing to the {\it direct} detection
for the spontaneous emission, while the latter type
has no such ``jump" onto the ground state $|g\ra$.
However, as expected, ensemble average of each type
of quantum trajectories (over 2000)
gives the same result of \Eq{ME-1}, as demonstrated in Fig.\ 2(b).

\vspace{0.2cm}
{\flushleft\it Summary and discussions}.---
We have constructed a scheme to generalize the QT theory
from Markovian to non-Markovian environments.
Taking the specific model of Lorentzian SDF,
we revealed a perfect scaling property
between the spectral bandwidth and the measurement frequency.
Our result bridges the gap between the existing QT
and the quantum Zeno effect by rendering them as two extremes.

While leaving the possible existence of scaling behavior
an open question for some non-Lorentzian SDFs,
the main conclusion above is valid in general.
Following the procedures in this work,
one can develop similar generalized QT theory
by numerically obtaining the $\bar{a}(\Delta t)$ in \Eq{rate-2},
rather than using the analytic Eqs.\ (\ref{scal}) and (\ref{rate-1}).
In Appendix C, we outline the solution scheme for arbitrary SDF.

Unlike the Markovian counterpart, ensemble average of the proposed QTs
does {\it not} describe the {\it reduced} state
given by tracing the environment degrees of freedom
from the entire (system-plus-environment) unitary wavefunction.
Since the successive measurements in the QT
destroy the correlation (memory) effect
between different {\it free} evolutions,
the ensemble-average state also differs from the one
resulted from averaging the nMSSE discussed in literature
\cite{Str98,Str99,Wis08,Dio08,Wis03}.
For non-Markovian environment, as pointed out by Wiseman {\it et al}
\cite{Wis08,Wis03}, the nMSSE is not consistent with any physical
quantum trajectories (i.e., having {\it no} physical interpretations).

For the relevance of the present work to possible experiment,
we may refer to the {\it partial collapse} quantum measurement
of the solid-state superconducting qubit
\cite{Katz06,Kor0707R,Kor0707P}.
The changed state reported there is conditioned on
a projective null-result at the final time $t$,
but not on ``continuous" or ``frequent" null-result
over the interval $(0,t)$.
For Markovian environment, both results are identical;
however, for non-Markovian case, this is not true.
Possible experiment may be guided by the formula \Eq{scal}
or (\ref{rate-1}), via the scaling variable $x$.
As alternative demonstration, one may perform a large-derivation
analysis on the emitted photons from driven atoms \cite{Gar10}.
From present work, we expect that
if altering the detection interval $\tau$
for the spontaneous emissions,
the statistics of the emitted photons
will be drastically different.
We would like to leave this interesting problem for future investigation.

\vspace{0.2cm}
{\flushleft\it Acknowledgements.}---
This work was supported by
the Beijing Natural Science Foundation under grant No.\ 1164014,
the Fundamental Research Funds for the Central Universities,
the NNSF of China under grants No.\ 91321106 \& 210100152,
and the State ``973" Project under grant No.\ 2012CB932704.

\appendix

\section{Zeno effect for superposition state}

{\flushleft Consider the superposition state}
$|\Psi(0)\ra=(\alpha_0|e\ra+\beta_0|g\ra)\otimes|{\rm vac}\ra$.
After small time $\tau$ the wave function becomes
\begin{align}
|\Psi (\tau)\rangle = [
\alpha_0(1-iH\tau -H^2\tau^2/2  +\cdots)|e\rangle
+ \beta_0 |g\rangle ] \otimes|{\rm vac}\ra  \,.
\end{align}
The null-result measurement in the environment implies
that the wave function is projected on the atomic subspace,
$|\Psi (\tau)\rangle\to \hat Q\,|\Psi (\tau)\rangle$,
where $\hat Q=\big(|e\rangle\langle e|+|g\rangle\langle g|\big)/{\cal N}$
and ${\cal N}$ is a normalization factor. Therefore
\begin{align}
|\Psi_1\rangle =\hat Q\,|\Psi (\tau)\rangle
= \Big[ \alpha_0\big(1-K\tau^2\big)|e\rangle
 + \beta_0|g\rangle \Big]/{\cal N}_1  \,,
\end{align}
where $K=\sum_r V_{r}^2$ and ${\cal N}_1^2=1-2\,\alpha_0^2\,K\tau^2$,
with $V_{r}$ the atom-environment (the $r$-th mode)
coupling amplitude.
After $n$ subsequent null-result measurements during time $t$,
with  $n=t/\tau$, we find
\begin{align}
|\Psi_n\rangle =\Big[ \alpha_0\big(1-K\tau^2\big)^n|e\rangle
+ \beta_0|g\rangle \Big]/{\cal N}_n \,,
\end{align}
where ${\cal N}_n=\sqrt{1-2\, n\, \alpha_0^2 K\,\tau^2}$.
Thus in the limit $\tau\to 0$ and $t$=const,
we obtain the result Eq.\ (2) in the main text ,
$|\Psi_n\rangle \to |\Psi (0)\rangle$.


\section{Solution for spontaneous emission}

{\flushleft Substituting Eq.\ (4) in the main text}
into the Schr\"odinger equation,
$i\partial_t |\Psi(t)\rangle =H|\Psi(t)\rangle$
and performing the Laplace transformation,
$\tilde f(\omega) =\int_0^\infty f(t)\exp (i \omega t)dt$,
we obtain the following system of algebraic equations:
\begin{subequations}
\label{eqs}
\begin{align}
&(\omega-E_e)\tilde \alpha (\omega)-\sum_r V_{r}\tilde c_r(\omega)
=i\alpha_0 \,,
\label{eqs1}\\
&[\omega-(E_g+\omega_r)]\tilde c_r(\omega)-V^*_{r}\tilde \alpha(\omega)=0 \,.
\label{eqs2}
\end{align}
\end{subequations}
The r.h.s. of the first equation reflects the initial condition.
Substituting $\tilde{c}_r(\omega)$ from Eq.~(\ref{eqs2})
into Eq.~(\ref{eqs1}), we obtain
\begin{align}
&(\omega-E_{e})\tilde \alpha(\omega)
-{\cal F}(\omega) \tilde \alpha(\omega )=i\alpha_0 \, ,
\label{a2}
\end{align}
where
\begin{equation}  \label{eq:6}
\mathcal{F}(\omega)=
\int\frac{D(\omega_r)}{\omega-(E_g+\omega_r)}\, d\omega_r  \,.
\end{equation}
Rather than the wide-band limit for the ``Markovian'' reservoir,
in this work we consider a finite-band spectrum
by taking the spectral density function (SDF)
$D(\omega_r)$ in the Lorentzian form,
\begin{align}
D(\omega_r)\equiv \sum_{r'}|V_{r'}|^2 \delta(\omega_r-\omega_{r'})
\to D_0\Lambda^2/[ (\omega_r-\omega_0)^2+\Lambda^2]\, ,
\label{lor}
\end{align}
with $\omega_0$ the spectral center,
$D_0$ the spectral height,
and $\Lambda$ the spectral width.
We obtain then
\begin{align}
\mathcal{F}(\omega )
={\Lambda\Gamma/2 \over (\omega-\omega_0-E_g)+i\Lambda},
~~{\rm where}~~
\Gamma=2\pi D_0 \,.
\end{align}
Substituting this result into Eq.~(\ref{a2}),
we find the amplitude $\tilde \alpha(\omega )$.
The time-dependent amplitude is obtained via the inverse Laplace transform,
$\alpha(t)=\int_{-\infty}^\infty \tilde \alpha(\omega)
e^{-i\omega t}d\omega /(2\pi)$.
Then, we obtain $\alpha(t)=a(t)\alpha_0$,
with an explicit expression of $a(t)$ given by Eq.\ (5) in the main text.

\section{ Solution scheme for non-Lorentzian SDF}

{\flushleft For Lorentzian SDF},
as shown above, we can first solve
\Eq{a2} in frequency domain, then obtain the analytic
solution of $\alpha(t)$ by means of inverse-Laplace transformation.
However, for arbitrary SDF $D(\omega_r)$,
this strategy does not work.
Instead, we can solve \Eq{a2} for $\alpha(t)$
numerically (and directly) in time domain.
For this purpose, an inverse-Laplace transformation
to \Eq{a2} yields
\begin{align}
i \dot{\alpha}(t)=E_e\,\alpha(t)+\int_{0}^{t} dt' \, F(t-t') \, \alpha(t') \,,
\label{a-t}
\end{align}
where
\bea\label{F-t}
F(t-t')&=&\int^{\infty}_{-\infty} \frac{d\omega}{2\pi} \,
e^{-i\omega (t-t')} {\cal F}(\omega) \nl
&=& -i\, \int d\omega_r D(\omega_r)\, e^{-i(\omega_r+E_g)(t-t')} \,.
\eea
Here we have employed the well known convolution formula
in Laplace transformation, and the following result
related to inverse-Laplace transformation
\bea
\int^{\infty}_{-\infty} \frac{d\omega}{2\pi}\, e^{-i\omega t}
[\omega-(\omega_r+E_g)]^{-1} = -i\,e^{-i(\omega_r+E_g)t} \, .
\eea
In practice, for a given SDF $D(\omega_r)$,
one can first carry out $F(t-t')$ in advance, via \Eq{F-t};
then, numerically integrate \Eq{a-t} to obtain $a(t)$.
With this result at hand, it is straightforward
to develop the generalized QT theory,
by numerically generating the $\bar{a}(\Delta t)$ in \Eq{rate-2},
rather than using the analytic Eqs.\ (\ref{scal}) and (\ref{rate-1}).
We have examined this numerical scheme
on the Lorentzian SDF
and found excellent agreement with the analytic solution.
The same success can be anticipated
when applying to arbitrary non-Lorentzian SDFs.

\end{document}